\documentclass[debug]{rmaa}

\usepackage{paralist}

\usepackage{psfrag,color}

\usepackage[latin1]{inputenc}

\usepackage{lscape}
\long\def\symbolfootnote[#1]#2{\begingroup%
\def\thefootnote{\fnsymbol{footnote}}\footnote[#1]{#2}\endgroup} 

\newcommand{\kms}{km\,s$^{-1}$}

\title{The filamentary multi-polar \\
        planetary nebula NGC\,5189 \symbolfootnote[2]{\scriptsize Based on observations obtained with the Swope and du Pont telescopes at Las Campanas Observatory and archival material from the Gemini Observatory and NASA/IPAC Infrared Science Archive}  }

\author{
  L. Sabin,\altaffilmark{1} 
  R. V\'azquez,\altaffilmark{1}
  J. A. L\'opez\altaffilmark{1}
  Ma.T. Garc\'ia-D\'iaz\altaffilmark{1}\\
  and G. Ramos-Larios\altaffilmark{2}}

\altaffiltext{1}{Instituto de Astronom\'{\i}a, Universidad Nacional Aut\'onoma de M\'exico, 
Km. 103 Carretera Tijuana-Ensenada, 22860 Ensenada, B. C., Mexico}
\altaffiltext{2}{Instituto de Astonom{\'i}a y Meteorolog{\'i}a, Departamento de F{\'i}sica, CUCEI, Universidad de Guadalajara, Av. Vallarta 2602, C.P. 44130, Guadalajara, Jal., Mexico}

\shortauthor{Sabin et al.  }
\shorttitle{The multi-polar planetary nebula NGC\,5189}

\fulladdresses{
\item Sabin, V\'azquez, L\'opez \& Garc\'ia-D\'iaz: 
\item IA-UNAM, Apdo. Postal 877, 22830 Ensenada, B. C., M\'exico 
\item ({lsabin,vazquez,jal,tere@astrosen.unam.mx})
\item Ramos-Larios :
\item IAM-UG, Av. Vallarta 2602, C.P. 44130, Guadalajara, Jal., M{\'e}xico
\item ({gerardo@astro.iam.udg.mx})

}
 
\listofauthors{ Sabin L., V\'azquez R., L\'opez J. A., Garc\'ia-D\'iaz Ma.T. \& Ramos-Larios G.}
\indexauthor{Sabin, L.}
\indexauthor{V\'azquez, R.}
\indexauthor{L\'opez, J. A.}
\indexauthor{Garc\'ia-D\'iaz, Ma. T.}
\indexauthor{Ramos-Larios, G.}

\abstract{We present a set of optical and infrared images combined with long-slit, medium and high dispersion spectra of the southern planetary nebula (PN) NGC\,5189. The complex morphology of this PN is 
puzzling and has not been studied in detail so far. Our investigation reveals the presence of a new dense and cold infrared torus (alongside the optical one) which probably generated one of the two optically 
seen bipolar outflows and which might be responsible for the twisted appearance of the optical torus via an interaction process. The high-resolution MES-AAT spectra clearly show the presence of filamentary and knotty structures as well as three expanding bubbles. 
Our findings therefore suggest that NGC\,5189 is a quadrupolar nebula with multiple sets of symmetrical condensations in which the interaction of outflows has determined the complex morphology.}
\resumen{Presentamos un conjunto de im\'agenes \'opticas e infrarrojas combinadas con espectros de
rendija larga de mediana y alta dispersi\'on de la Nebulosa Planetaria (NP) del sur NGC\,5189. La
compleja morfolog\'{\i}a de esta NP es desconcertante y no hab\'{\i}a sido estudiada en detalle hasta
ahora. Nuestra investigaci\'on revela la presencia de un toroide denso y  
fr\'{\i}o, en el infrarrojo, el cual probablemente gener\'o uno de los dos flujos bipolares vistos 
en el \'optico y podr\'{\i}a, mediante un proceso de interacci\'on, ser tambi\'en responsable de la apariencia retorcida del toroide \'optico. Los espectros de alta resoluci\'on del MES-AAT muestran claramente la presencia de nudos y estructuras filamentosas, as\'{\i} como tres burbujas en expansi\'on. 
Nuestros hallazgos sugieren que NGC\,5189 es una NP cuadrupolar con varios conjuntos de condensaciones sim\'etricas
donde la interacci\'on de flujos determin\'o su compleja morfolog\'{\i}a.}

\addkeyword{ISM:planetary nebulae: individual (NGC\,5189)}

\begin{document}
\maketitle

\section{Introduction}

NGC\,5189 (PK 307-03$\arcdeg$1, PN G307.2$-03.4$, He 2$-94$) belongs to a
group of bipolar planetary nebulae that exhibit more than one bipolar
structure. Examples of objects in this class are NGC\,2440 \citep{Lopez1998}, NGC\,6302 \citep{Meaburn2008}, NGC\,6309 \citep{Vazquez2008}, NGC\,1514 \citep{Aryal2010} and NGC\,6644 \citep{Hsia2010}. These kind of objects represent particularly interesting cases of study as they show morphological signatures of episodic mass
outflows, most likely emerging from a precessing source. This type of
phenomenon is receiving increasing attention in the literature, and is
found not to be restricted to PNe with bipolar morphologies. The point symmetric IC\,4634 \citep{Guerrero2008} and the elliptical Fleming\,1 (PN G290.5+07.9) \citep{Lopez1993} are examples of objects that have been discussed under assumptions of
episodic, symmetric outflows with precessing symmetry axis. In the case of
elliptical PNe, instead of multiple lobes one finds twisted strings of
ansae and condensations delineating the presumed precession of the symmetry
axis. Theoretical models like those of \citet{Soker1989,Soker1994}, based on close binary systems combined with a precessing systemic axis describe one of the plausible scenarios under which such conditions may occur. The implication of magnetic fields should not be discarded as magnetohydrodynamic (MHD) models such as those by \citet{Segura1999} were able to reproduce the collimated outflows (or jets) observed by coupling magnetized winds and stellar rotation.

\citet{Phillips1983} have pointed out the interesting structure of NGC 5189 which they describe at first sight as ``highly chaotic''. In their electronographic narrow band images, they identified five pairs of condensations or ansae (low-ionization structures following \citealt{Goncalves2001}) symmetrically distributed with respect to the central star (CS) at different position angles. The authors infer the occurrence of multiple mass loss events with a change of orientation in time to explain the observed pattern. And to account for the precessing phenomenon, they suggest the presence of a binary system (sheltering the precessing central star) with a few days period. To our knowledge no binary companion has been detected so far in NGC\,5189.

\citet{Reay1984} used a wide field imaging Fabry-Perot system to obtain a bidimensional line profile map of NGC\,5189. By this mean they derived the kinematics of two potential pairs of ansae and their results indicate rather low-medium velocities. 
Indeed, the first pair, which corresponds to the smallest and closest (to the CS) condensations as defined by \citet{Phillips1983}, is described as an inner ring with an expansion velocity of 25.3\,{\kms} (radially, from the CS). The radial velocities of the second more distant pair of ansae to the CS following \citet{Phillips1983} have been derived as $\sim${$|8|$}\,{\kms}.

Despite its significant asymmetry, NGC\,5189 has not been extensively studied from a morphological point of view and the main information collected are gathered in Table~\ref{ngc5189data}. Also, in this paper we present unpublished optical imaging data, medium-- and high resolution spectra as well as newly analysed infrared images from the Midcourse Space Experiment  (\emph{MSX}; \citealt{Price01}) and the recent Wide-field Infrared Survey Explorer (\emph{WISE}; \citealt{Wright2010}) surveys. The combination of datasets will help us to analyse in detail the morphological structure of the PN.

\begin{table}[!t]\centering
  \setlength{\tabnotewidth}{0.5\columnwidth}
  \tablecols{3}
  \setlength{\tabcolsep}{2.8\tabcolsep}
  \caption{Physical properties of NGC\,5189 and its central star}\label{ngc5189data}
 \begin{tabular}{ll}
    \toprule
    Property & \multicolumn{1}{l}{Value}   \\
    \midrule
    Type PN                   & I$^{a}$, IIa$^{b}$  \\
    Distance (kpc)            & 0.88$^{c}$, 0.90$^{d}$, 0.55$^{e}$, 0.70$^{f}$, 0.54$^{l}$ \\
    $c$(H$\beta$)$^{\it Balmer}$& 0.65$^{a}$, 0.85$^{l}$\\
    $c$(H$\beta$)$^{\it Radio}$ & 0.56$^{a}$, 1.83$^{l}$\\
    $S_{\nu}$(5GHz) (mJy)     & 455$^{c}$, 507$^{l}$ \\
    $\Theta_{\rm diam}$ (\arcsec) & 140$^{c}$, $163.4\times108.2^{j}$ \\
    $T_{\rm e}$([O\,III]) ($10^{4}$\,K)     & 1.31$\pm$0.05$^{a}$\\
    $N_{\rm e}$([S\,II]) (cm$^{-3}$)     & 390$\pm$60$^{a}$, 200--900$^{i}$, 1000$^{l}$\\
    $T_{\rm eff}$ (K)             & 135$\times$10$^{3}${$^{k}$} \\
    Stellar Type              & [WC 2]$^{g}$, [WO I]$^{h,i}$ (O VI sequence) \\
    log ($L_{\star}$/L$_{\sun}$)    & 2.9$\pm$0.5$^{k}$\\
     \bottomrule
    \end{tabular}
     \begin{minipage}[!t]{10cm}
   References: $^{a}$\citet{Kingsburgh1994}, $^{b}$\citet{Quireza2007}, $^{c}$\citet{Zhang1995}, 
   $^{d}$\citet{Tajitsu1998}, $^{e}$\citet{Phillips2004}, $^{f}$\citet{Maciel1994}, $^{g}$\citet{Mendez1991}, 
   $^{h}$\citet{Miszalski2009}, $^{i}$\citet{Polcaro1997}, $^{j}$\citet{Tylenda2003}, $^{k}$\citet{Althaus2010}, 
   $^{l}$\citet{Hua1998}.
\end{minipage}
\end{table}

\begin{figure}[!t]
{\includegraphics[width=\textwidth]{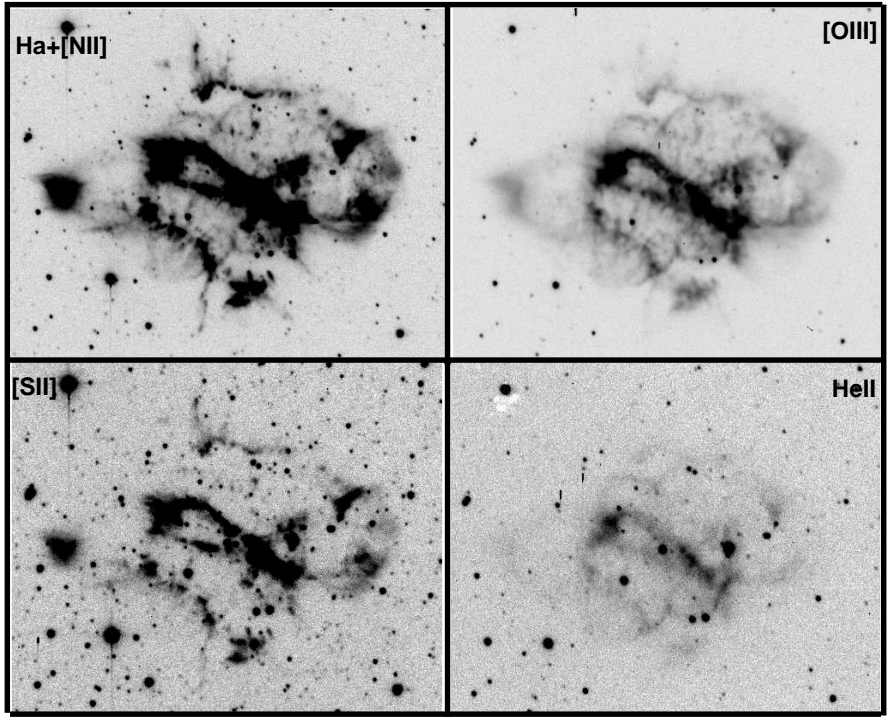}}
\caption{\label{ngc5189im}Logarithmic scaled images of NGC\,5189 in H$\alpha$+[N\,II], [O\,III], [S\,II] and He II taken with the 1.0 m Swope telescope at LCO. North is up and east is left in a field of view of 
$\simeq3\farcm7\times3\farcm0$.}
\end{figure}

\begin{figure}[!t]
\begin{center}
{\includegraphics[width=\textwidth]{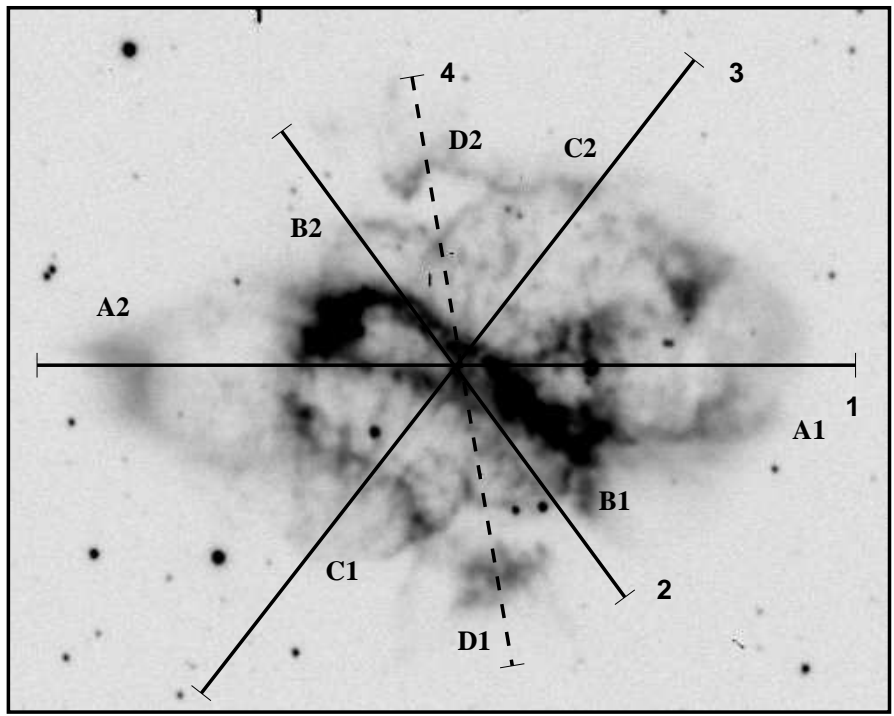}}
\caption{\label{ngc5189out}Outflows location in the PN NGC\,5189. The dash lines indicates a possible fourth ejection axis }
\end{center}
\end{figure}

\begin{figure}[!t]
{\includegraphics[width=\textwidth]{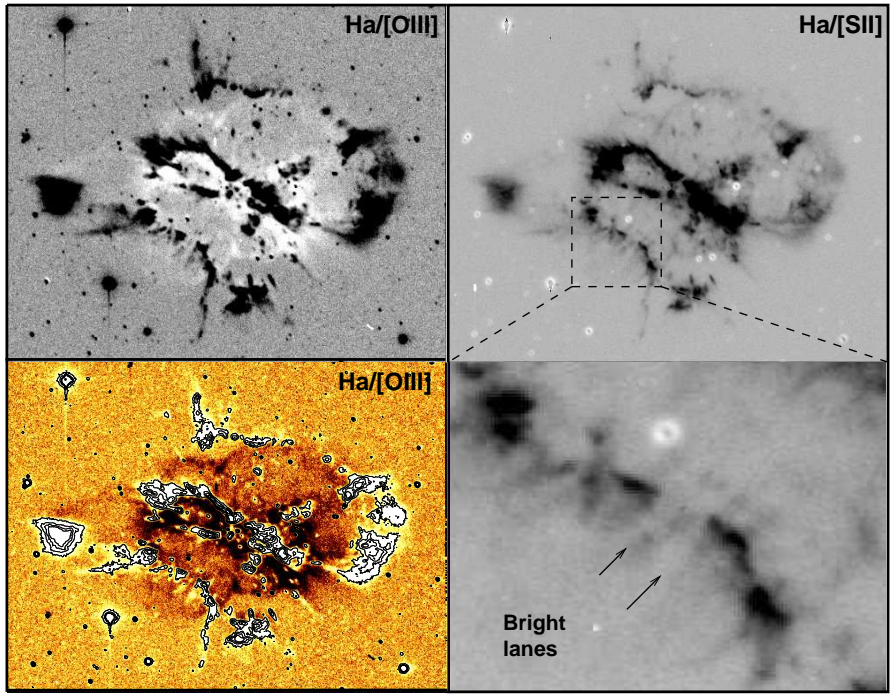}}
\caption{\label{ngc5189ima}Top: Image ratio using the H$\alpha$, [O\,III] and [S\,II] filters and a closer view of the knots alignment in the bottom right frame (zoom of top right frame). In each image ratio, black regions refer to high ratios, white regions refer to low ratios. Bottom left: Highlighted contours of the high excitation zones (condensations) in the H$\alpha$/[O\,III] image. The [O\,III] emission is mainly concentrated inside the H$\alpha$+[N\,II] nested bubble. (Note: For the color version of this Figure see the digital version of the journal). Bottom right: Dark lanes which appear as bright paths in the inverted image.}
\end{figure}

\begin{figure}[!t]
\begin{center}
{\includegraphics[width=\textwidth]{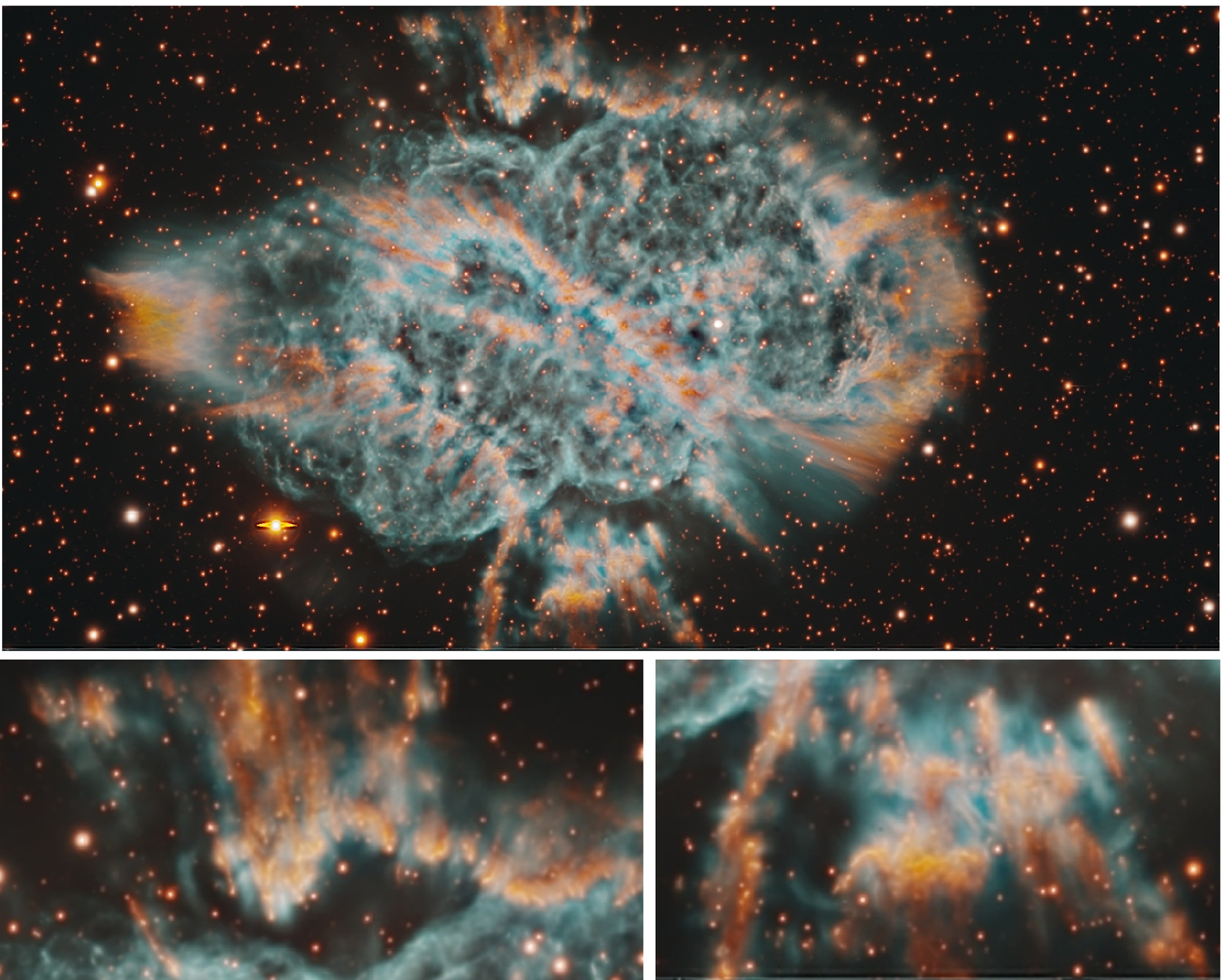}}
\caption{\label{ngc5189gemini} Image of NGC\,5189 processed with PixInsight underlying with unprecedented depth the detailed structures of the PN. The northern and southern caps (left and right bottom panels respectively) are maily composed of cometary knots. Credit: Juan Conejero with original raw data from the Gemini Science Archive (H$\alpha$: orange, [O\,III]: blue and [S\,II]: red). For the color version of this Figure see the digital version of the journal.}
\end{center}
\end{figure}

\section{Observations and Results}

\subsection{Optical Imaging} 

A set of images of NGC\,5189 were obtained at Las Campanas Observatory (LCO, Chile) on the 1.0 m Swope telescope in May 1990, with a Texas Instruments (TI\#1) CCD detector. The effective scale is 0.435 arcsec/pixel corresponding to a nominal field of 5.8 square arcmin. The interference filters were centred on 
H$\alpha$ $\lambda6563$ ($\Delta\lambda\simeq78$\AA), [O\,III] $\lambda5007$ 
($\Delta\lambda\simeq70$\AA), [S\,II] $\lambda6724$ ($\Delta\lambda\simeq76$\AA) and He\,II 4686 
($\Delta\lambda\simeq70$\AA) with exposure times of 600\,s (H$\alpha$ and [O\,III]), 900\,s and 1200\,s, respectively. The images, presented in Fig.~\ref{ngc5189im}, were trimmed to $800\times800$ pixels and reduced with the standard {\sc iraf}\footnote{ IRAF is distributed by the National Optical Astronomy Observatory, which is operated by the Association of Universities for Research in Astronomy (AURA) under cooperative agreement with the National Science Foundation} methods.

The [O\,III] frame (and to a lesser extent the He\,II one) in Fig.~\ref{ngc5189im} perfectly delineates the highly excited multiple outflows remarking the complex morphology while the H$\alpha$+[N\,II] and [S\,II] frames underline the central part and borders of NGC\,5189. 
Fig.~\ref{ngc5189out} presents the three likely outflows seen in NGC\,5189 (well illustrated in the [O\,III] image) and numbered 1,  2, and 3 orientated at the respective position angles (PA) of 90$\arcdeg$, 37$\arcdeg$ and 141$\arcdeg$ and the respective associated pairs of collimated outlows A1--A2, B1--B2 and C1--C2. The existence of a fourth outflow (numbered 4 on Fig.~\ref{ngc5189out}) at PA 10$\arcdeg$ is less clear as only the `polar' caps of the probable ejections are seen (D1 and D2). This case will be discussed later in the article. All the bipolar structures are emerging from the same geometric center related to the central star. The condensations or ansae defined by \citet{Phillips1983} which include the opposite structures named B1, B2, and C by \citet{Reay1984} are clearly identified in our images (and particularly in the image ratio H$\alpha$(+[N\,II])/[O\,III], Fig.~\ref{ngc5189ima}) and correspond to the tips of the axes 1 and 4 (A2, D1, and D2). The H$\alpha$(+[N\,II])/[O\,III] frame which traces the excitation conditions, also shows that the high excitation structures are mainly related to the central toroidal-like belt and the outer edges of NGC\,5189. 
It is also interesting to notice from the image ratio in Fig.~\ref{ngc5189ima} that the [O\,III] emission is globally enclosed in this H$\alpha$+[N\,II] `nested bubble' and seems to form a bipolar structure whose waist is coincident with the axis 4 and whose long axis is coincident with the axis 1. The [S\,II] emission is distributed similarly to that of H$\alpha$+[N\,II] and from 
Fig.~\ref{ngc5189ima} (top and bottom right), it appears that the string of knots forming the outer edges (the opposite North-West and South-East rims), may in fact reflect a discontinuity in the high excitation outer shell of the PN. The poly-symmetric structure of NGC\,5189 is clearly related to its dynamic evolution. A signature of this dynamical process are the dark lanes well seen in between knots of the South-East outer rim in the non-inverted H$\alpha$ and [S\,II] images.

All images in Fig.~\ref{ngc5189im} clearly show the presence of a knotty central toroidal structure with a major axis size of about 78\arcsec and a PA of $\sim$ 62$\arcdeg$. This structure seems flatten, tending more towards a disk description, with radial filaments emerging from its outer border. This torus/disk seems also to be slightly twisted.  
Several collimated cometary knots are located in NGC\,5189 and are well seen in Fig.~\ref{ngc5189gemini}. Those knots suggest the interaction of the stellar wind and photoionizing flux with the nebular material. 

 \subsection{IR imaging}

\begin{figure}
\begin{center}
{\includegraphics[height=3.7cm]{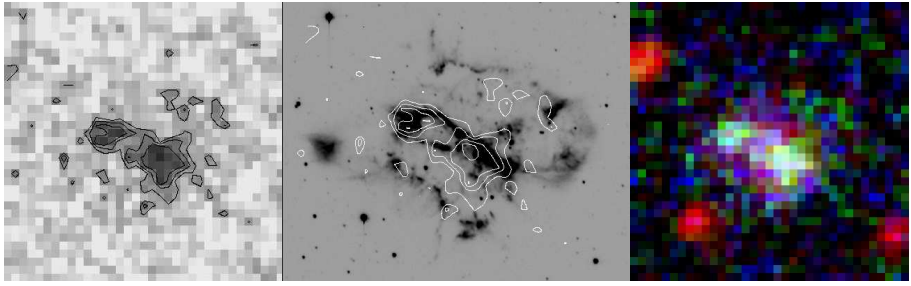}}
\caption{\label{MSX}From left to right (North is on the top and East on the left): (i) MSX 3.5 arcmin$^2$ IR image and contours of NGC\,5189 in D band (14.65-$\micron$). (ii) Contours of the MSX Band D emission overlaid to an optical image of NGC\,5189. The IR emission matches the toroidal structure around the central star. (iii) Composite image of three MSX bands (A: red, D: green, and E: blue). For the color version of this figure see the digital version of the journal.}
\end{center}
\end{figure}

The dust and molecular content of PNe has often been used as morphological tracer in the infrared wavebands (e.g. \citealt{Phillips08,Ressler2010,Ramos-Larios2008}). We present Near-, Mid-Infrared (NIR and MIR) and H$_{2}$ images of NGC\,5189. In Fig.~\ref{MSX} we show the D-band (14.65-$\micron$) image of the MIR \emph{MSX} as well as a composite image of the data bands A, D, and E at the respective wavelengths 8.28-$\micron$, 14.65-$\micron$ and 21.34-$\micron$. The respective fluxes being 0.216 Jy at 8.28-$\micron$, 1.029 Jy at 14.65-$\micron$, and 1.635 Jy at 21.3-$\micron$, with the respective quality flux of 3, 2 and 1 ({\it i.e.} good, fair and limit; \citealt{Egan2003}). The spatial resolution of $\simeq18\farcs3$ of the Spatial Infrared Imaging Telescope which gathered the MSX data, allows us to distinguish a dusty structure which is found to be coincident with the torus and some parts of the outer borders of the PN when compared to an optical image. From the \emph{MSX} data, mainly the D band, it is clear that the dust is not equally distributed with more material in the southern side of the toroid than its northern side. But this effect could be due to the projection of the toroid.

\emph{WISE} images, from the preliminary catalogue and image atlas, were retrieved from the NASA/IPAC Infrared Science Archive (IRSA). \emph{WISE} is a NASA Explorer mission that survey the entire sky at wavelengths of 3.4, 4.6, 12, and 22-$\mu$m (W1 through W4, respectively). An advantage of \emph{WISE} is its angular resolution better than that of \emph{MSX} with 6$\farcs$1, 6$\farcs$4, 6$\farcs$5 and 12$\farcs$0 at 3.4, 4.6, 12, and 22-$\mu$m respectively.

\begin{figure}[!t]
\begin{center}
{\includegraphics[width=\textwidth]{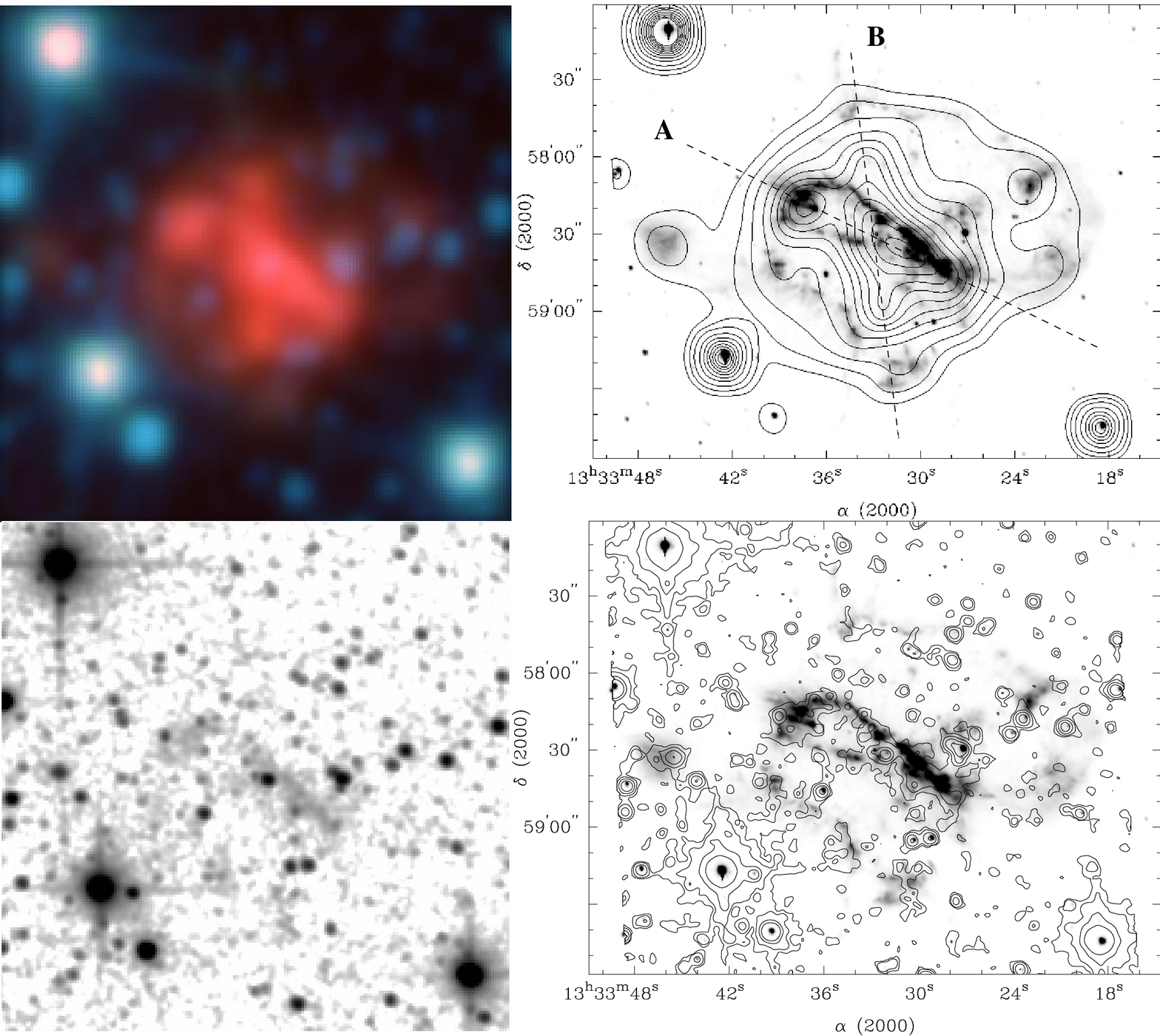}}
\caption{\label{WiseH2} Infrared imaging analysis of NGC\,5189. Top Left and Right: Composite image of NGC\,5189 using the 3.4-$\mu$m (blue), 4.6-$\mu$m (green) and 12-$\mu$m (red) and highlighting the central {\it X}-shaped structure. The contours of the 12-$\mu$m emission are overlaid on a H$\alpha$+[N\,II] image in the next image. Bottom Left and Right: H$_{2}$ map of NGC\,5189 obtained with the 2MASS $H$ and $K_s$ broad bands. The contours overlaid on a H$\alpha$+[N\,II] indicate that the main optical torus is also emitting in molecular hydrogen. Other external structures such as the cometary knots and some parts of the edges are also emitting in this band. For the color version of this figure see the digital version of the journal.}
\end{center}
\end{figure}

Fig.~\ref{WiseH2}-Top Left shows a 3-colors image of NGC\,5189 using the 3.4-$\mu$m (blue), 4.6-$\mu$m (green) and 12-$\mu$m (red) [close to the MSX D band] where we immediately notice a bright {\it X}-shaped structure which also seems to display much fainter ansae. The peculiar central morphology is well seen at the largest wavelengths 
{\it i.e.} in the W3 and W4 wavebands (12 and 22-$\mu$m where the derived magnitudes are the highest with 5.7 mag and 2.0 mag respectively) which allow us to measure an opening angle of $\simeq59\arcdeg$ between both arms which also have nearly the same length ($\simeq70\arcsec$ for the arm A and $\simeq73\arcsec$ for the arm B). By comparing the WISE and optical morphologies we can appreciate the dusty nature of NGC\,5189 with the 12-$\mu$m (and 22-$\mu$m) emission perfectly covering the whole nebula and coinciding with its optical edges. While the arm A is fully coincident with the optical torus, the arm B does not show any obvious optical counterpart i.e. there is no other visible optical torus. 

Finally, in order to estimate the distribution of molecular material in NGC\,5189 we have retrieved the $H$ and $K_s$ data from the 2MASS catalogue and followed the prescription by \citet{Ramos-Larios2006} to infer the H$_{2}$ distribution in the nebula. The resultant map is shown in Fig \ref{WiseH2}-Bottom (left and right) that we compare to the optical (H$\alpha$+[N\,II]) image. The results indicate that the molecular emission is mainly concentrated in the optical torus and there is some indication that H$_{2}$ is also found in the nebula's edges and particularly at the northern and southern cometary knots locations. The latter is in agreement with the molecular H$_{2}$ content generally found in this kind of structures.

\subsection{Kinematics}

\begin{figure}[!t]
\begin{center}
\hspace{-0.7cm}
{\includegraphics[height=9cm]{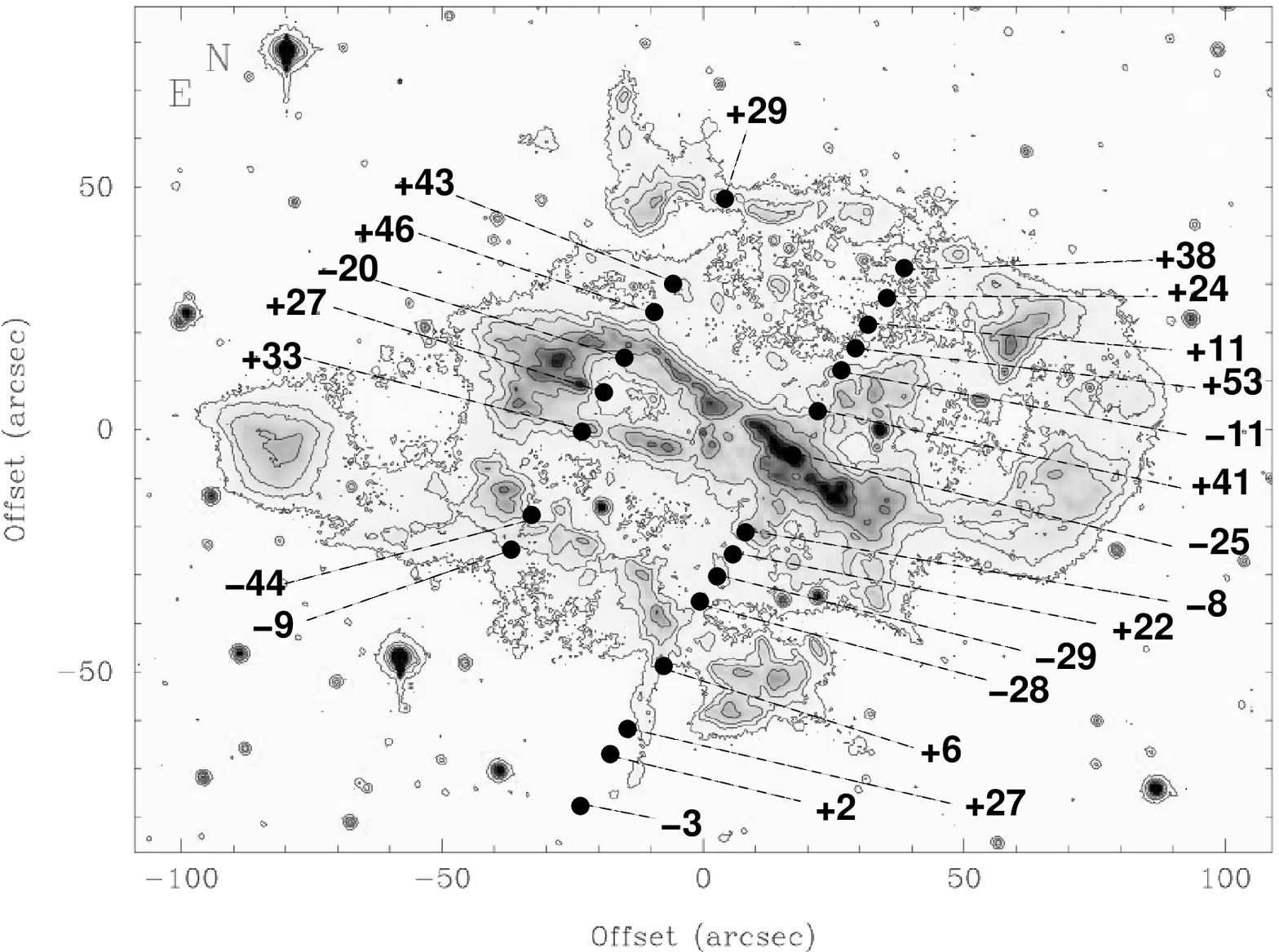}}
\caption{\label{ngc5189kine}Radial velocities along the LCO slits relative to the systemic velocity with an accuracy of $\simeq$4 {\kms}. }
\end{center}
\end{figure}

Longslit spectra were obtained with the modular spectrometer on the 2.5-m
du Pont telescope at LCO in June 1990. The spectrometer was used with its
200-mm camera combined with a CRAF CCD $1024\times1024$ pixels, with 
12-$\mu$m/pixel size. A 1200 lines/mm grating was used to cover the spectral
range from $\simeq\lambda\lambda6290$--6805\,\AA. The instrumental
combination yields a spectral resolution of $\simeq2$\,FWHM in pixels (linear dispersion
of 0.479\,\AA/pixel) and a spatial resolution along the slit of 0.292\,arcsec/pixel. 
A slit width of 1{\arcsec} and 149{\arcsec} long was set. The spectra
were flat-fielded and wavelength calibrated against separate exposures of a
Quartz and Neon arc, respectively. Two exposures were obtained, one of 1200\,s
along SE-NW axes and the other one of 1000\,s, at the same position angle 
PA=329$\arcdeg$, but shifted 36\farcs6 to the West (Fig.~\ref{ngc5189kine}).

\begin{figure}[!h]
\begin{center}
{\includegraphics[height=9cm]{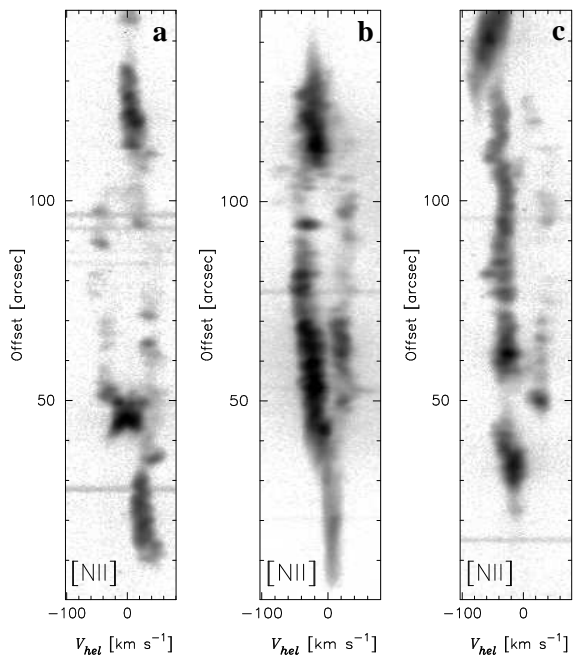}}
\caption{\label{AATNII} AAT-MES [N\,II] $\lambda6583.454$\,{\AA} slits details underlying the filamentary structure of NGC\,5189. }
\end{center}
\end{figure}

\begin{figure}[!h]
\begin{center}
{\includegraphics[height=\textwidth,angle=-90]{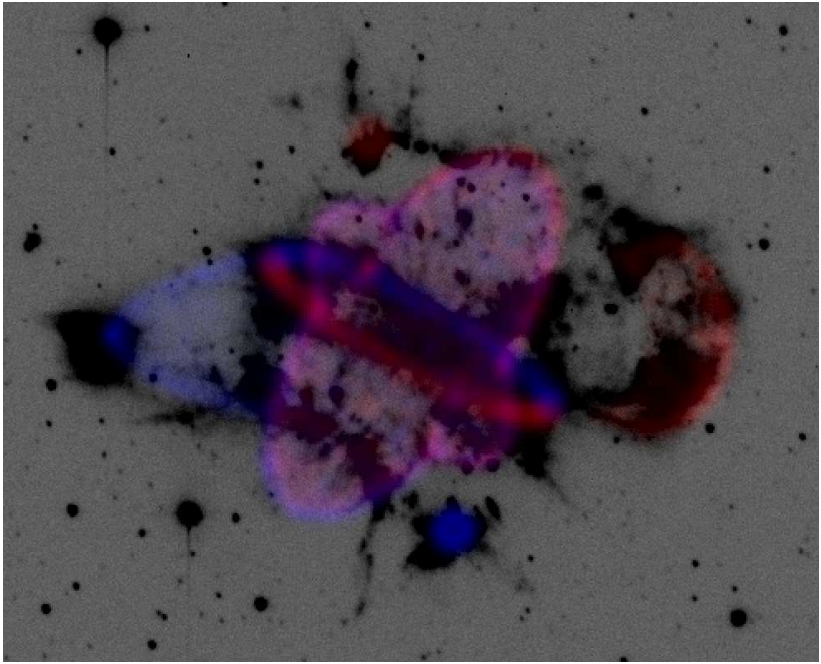}}
\caption{\label{Shape} Global kinemetical evolution of NGC\,5189 based on the velocity measurements from the LCO and MES spectra. The blue components are blue-shifted and the red ones are red-shifted. The ``model'' was realised using the morpho-kinematical tool Shape \citep{Steffen2006}. For the color version of this Figure see the digital version of the journal.}
\end{center}
\end{figure}

\begin{table}[!h]
\centering
   \caption{\small Velocity points from the AAT-MES spectral data calculated with a LSR corrected systemic velocity of $-13.3$\,{\kms} and the spectral emission line [N\,II] $\lambda6583.454$\,\AA. The error on the measurement is $\simeq1$\,{\kms} }
   \label{MESvel}
 \begin{tabular}{|c|c||c|c||c|c|}
    \toprule
 Slit A & $V_{\rm lsr}$& Slit B &   $V_{\rm lsr}$& Slit C &  $V_{\rm lsr}$ \\
 \hline
A1  &  +27   & B1   & +32   &C1  & $-28$ \\
A2  &  +40   & B1'  & $-34$   &C2  & $-27$ \\
A2' &  $-26$   & B2   & $-20$   &C2' & +13   \\
AA  &  +11   & B2'  & +27   &CA  & $-35$ \\
AB  &  $-5$    & BA   & $-12$   &CB  & $-46$ \\
AB' &  +14   & BB   & $-13$   &CC  & $-10$ \\
AC  &  +52   &      &       &    &       \\
AD  &  +26   &      &       &    &       \\
  \bottomrule   
      \end{tabular}
     \begin{minipage}[!t]{12cm}
 The data points with (') indicate a zone of expansion and correspond to the velocities measured using the faintest part of this expansion .
\end{minipage}
\end{table}

\begin{figure}[!h]
\begin{center}
{\includegraphics[width=\textwidth]{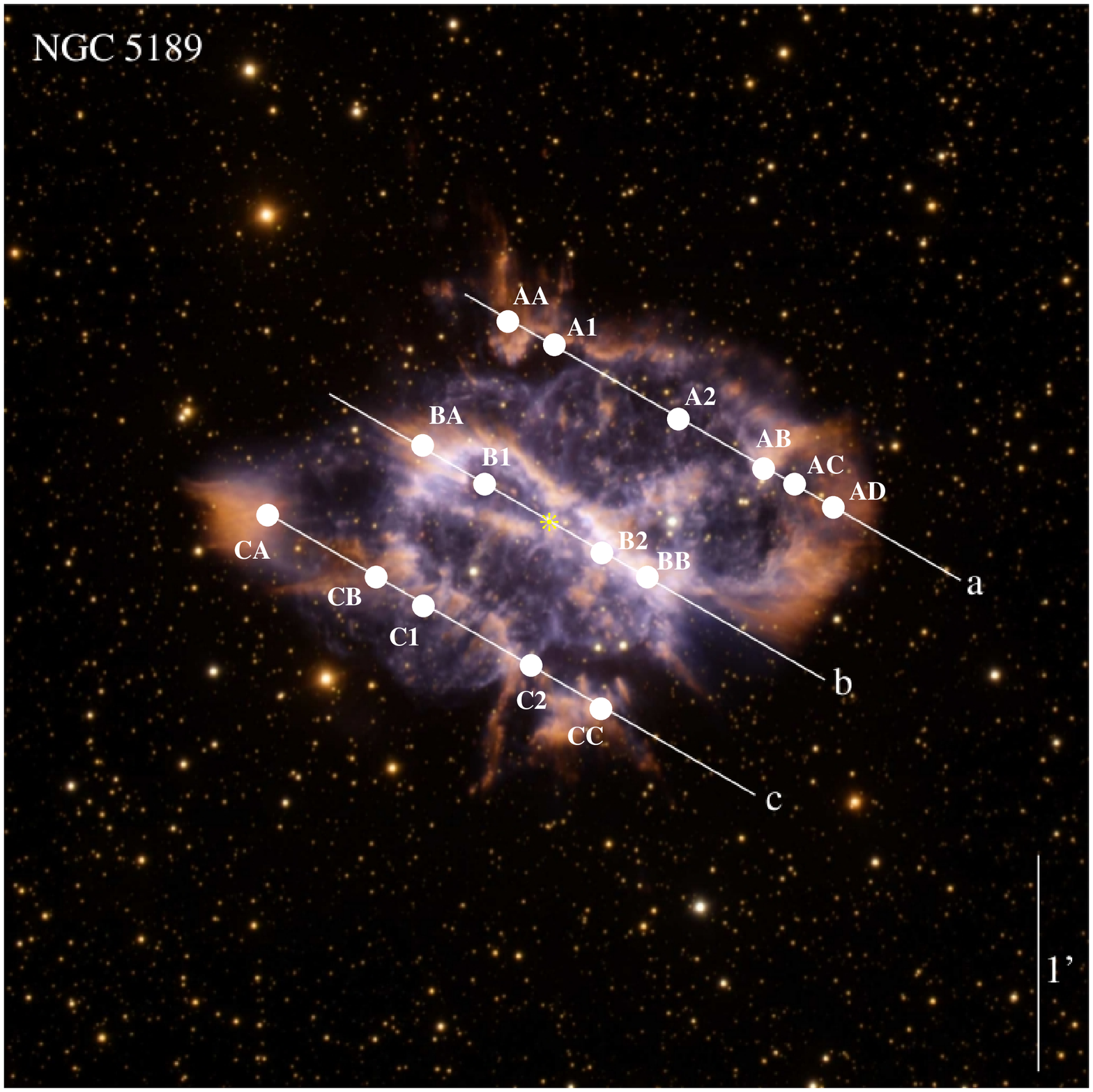}}
\caption{\label{AATslit2}AAT-MES slits positions superimposed to a $5\arcmin\times5\arcmin$ composite GMOS image (H-$\alpha$: orange, [O\,III]: blue and [S\,II]: red) from the Gemini South Observatory. North is up, East is left. The white dots indicate the areas where were derived the LSR corrected velocities: with A1, A2, B1, B2, C1 and C2 roughly corresponding to coinciding zones between the LCO and MES slits and the remaining points indicating the velocities of structures of interest such as the  cometary knots ({\it cf} text and Table \ref{MESvel}). The yellow star shows the central star location. For the color version of this figure see the digital version of the journal.}
\end{center}
\end{figure}

The medium resolution spectra allow us to visually distinguish variations in shape along each spectral line and velocity patterns. With a spectral dispersion of 0.479\,\AA/pixel over the spectral range yielding a resolution of 21.8\,{\kms}/pixel, we were able to estimate the radial velocities along NGC\,5189 at the two slits positions using the bright [N\,II] $\lambda6583$ line, adopting a rest wavelength 
$\lambda_{\rm lab}=6583.454$\,\AA. The accuracy on the wavelength calibration is about 4\,{\kms}. As none of the slits passes through the location of the central star of the nebula, we cannot directly estimate the systemic velocity. However this information was provided trough our high resolution spectroscopy data (see below) where we measured a systemic velocity $V_{\rm lsr}$ of $-13.3\pm1$\,{\kms}.
Fig.~\ref{ngc5189kine} shows the velocities at different points of the PN and although we only have the velocity pattern along the two slits, we can give an overall description of the kinematics considering NGC\,5189 as a large butterfly type (underlined by the external contours). Indeed, in adequacy with Fig.~\ref{ngc5189kine} it therefore appears that globally the northern part of the ``optical'' torus is blue-shifted (with negative values of -20 and -25 \kms) while its southern part is red-shifted (with mainly the positive velocity +33 \kms), giving us the orientation of inclination (Fig.~\ref{Shape}).  And as one would expect (by pure geometrical effect in an expanding axisymmetric nebula), the southern lobe of the butterfly envelope is mainly blue shifted (Fig.~\ref{ngc5189kine} indicates mainly negative velocities down to -44 \kms) while its northern counterpart is red shifted (with a majority of positive values up to +53 \kms. This large scale kinematics of the nebula is in agreement with the results obtained by \citet{Reay1984}.

In addition to the LCO spectra, long-slit high resolution spectra better suited for kinematic work, have been obtained with the Manchester Echelle Spectrometer (MES; \citealt{Meaburn1984}) on the 3.9-m Anglo-Australian telescope (AAT). The observations, performed on 1997 April 14 with a Tektronix CCD with 
$1024\times1024$ rows of 24-$\mu$m square pixels, were obtained with a 100\,{\AA} wide filter containing 
H$\alpha$ and the [N\,II] $\lambda\lambda6548$, 6584{\AA} emission lines. The spectral dispersion of 
$\simeq0.05$\,\AA/pixel over the spectral range yields to a resolution of 11\,{\kms}/pixel through a slit of width 150-$\micron$. The total slit coverage is composed of 3 exposures of 1200\,s each, all at the position angle PA=61$\arcdeg$ at opposite location respective to the central star and the torus long axis 
(Fig.~\ref{AATslit2}). The CCD was binned by two in the spatial direction, giving $512\times0.32$\,\arcsec/pixel. These spectra are available in the {\it SPM Kinematic Catalogue of Galactic Planetary Nebulae}\footnote{http://kincatpn.astrosen.unam.mx} \citep{Lopez2011}. 
The structure of the line profiles, presented in Fig.~\ref{AATNII}, is outstanding and reflects the highly filamentary and knotty structure of NGC\,5189. 
Similarly to the LCO spectra, the different velocities obtained with the MES spectra were derived using the bright [N\,II] $\lambda6584$\,{\AA} emission line. We identified a systemic velocity $V_{\rm lsr}$ of $-13.3\pm1$\,{\kms} 
($V_{\rm hel}=-6.8\pm1$\,{\kms}) which is comparable to the velocity derived by \citet{Schneider1983} within their error bar.

Also, we obtained the following maximum expansion velocities: $\Delta V=35$\,{\kms} in the North-West lobe (slit a), $\Delta V=33$\,{\kms} in the central region (slit b) and $\Delta V=44$\,{\kms} in the South-East lobe (slit c). Interestingly, those velocities do not show extreme values as would suggest the complex morphology. 

The MES--AAT spectra are positioned at $\simeq92\arcdeg$  with respect to that of the LCO and provide the same global kinematic information related to the main structures evolution as it can be seen in 
Fig.~\ref{Shape}.
Fig.~\ref{AATslit2} presents the detailed kinematics of several zones within NGC\,5189 ({\it cf} Table \ref{MESvel}). The velocities of the structures belonging to the central torus at both opposite sides (points BA and BB) evolve at pretty much the same rate with relative $V_{\rm lsr}$ values of $-12$ and $-13$\,{\kms} respectively, similar to the systemic velocity. The torus is moving homogeneously towards us and therefore would not have the ``warped'' characteristics where one would expect to see an equally strong red-shifted side. The higher velocities at the expanding points B1 ($+32$\,{\kms} and $-34$\,{\kms} ) and B2 ($-20$\,{\kms}, $+27$\,{\kms}) could be the result of the dynamic motions of structures which come across the torus. The cometary knots in the northern and southern parts of the nebulae (points AA and CC respectively aligned with axis 4 in Fig.~\ref{ngc5189out}) show roughly the same velocities with +11 and $-10$ respectively. Finally, the zones of NGC\,5189 located on the [N\,II] rich butterfly edges (see 
Fig.~\ref{ngc5189ima}-Top Left) have been studied (points CA, CB and C1 on the South-East and A1, AC and AD on the North-West) and they globally show the same larger velocities. Thus, the mean velocity of the components of the ``southern lobe'' is $\simeq-36$\,{\kms} while the mean velocity of the components of the ``northern lobe'' is $\simeq+35$\,{\kms}.

\section{Discussion and Conclusion}

A detailed study of the seemingly complex structure of NGC\,5189 has revealed a poly-polar morphology. The main findings are discussed below.
\begin{itemize}
\item
This work highlights the presence of a new dense and cold infrared torus whose coincidence with the waist seen in the axis 4 in Fig.~\ref{ngc5189out} indicates that it gave birth to a bipolar outflow which coincides with the nested bubble of [O\,III] emission. The interaction of this newly discovered IR torus and the optical one (which are related to the same geometric center) might explain the warped morphology of the latter. The twisted torus hypothesis is ruled out by the kinematics measurements.
\item
The MES-AAT spectra indicate the presence of filamentary and knotty structures as well as the expansion of three bubbles with the following velocities: $\Delta V=35$\,{\kms} in the North-West lobe (slit a), 
$\Delta V=33$\,{\kms} in the central region (slit b) and $\Delta V=44$\,{\kms} in the South-East lobe (slit c). 
\item
The high resolution spectroscopy data also show that the largest velocities are found at both extremes of the nebulae with $-35$\,{\kms} and $-46$\,{\kms} for the points CA and CB respectively on the eastern side of the PN (Fig.~\ref{AATslit2}), and +26\,{\kms} and +52\,{\kms} for the opposite points AD and AC on the western side. The similar velocity of the diametrically opposed structures named AA and CC with +11\,{\kms} and 
$-10$\,{\kms} seems to indicate that both are related.
\end{itemize}

We therefore suggest that NGC\,5189 is composed of two symmetry axis each of them composed of a torus associated to a bipolar outflow. This corresponds to the IR torus combined to the [O\,III] bipolar emission which encloses the outflows B1--B2, C1--C2 and part of A1--A2 (see Fig.~\ref{ngc5189out} and \ref{ngc5189gemini}); and the optical torus combined to the bright outer filamentary/knotty butterfly structure whose contours are well seen in the H$\alpha$/[O\,III] image ratio in Fig.~\ref{ngc5189ima}. Any other lobes would be the result of the expansion or dynamical motions of the existing nebular material in low-density or empty spaces. 

The combination of the imaging and spectroscopic data allowed us to make a first step in disentangling and (re-)defining NGC\,5189 exact morphology. Indeed, the ``chaotic'' aspect of NGC\,5189 is the consequence of the encounter between two sets of bipolar outflows. The boundary between the different outflows is however not well determined with our kinematic data, although we can see it in the morphology, nor is the process leading to such evolution of the planetary nebula. We therefore need a more complete kinematical mapping ({\it e.g.} to derive the inclination angles and the kinematical ages of each structures) to have a better understanding of the formation of NGC\,5189.

\section*{Acknowledgements}
This work is supported by PAPIIT-UNAM grant IN109509 (Mexico). LS acknowledges a Post-Doctoral fellowship from UNAM.
Based on observations obtained at the Gemini Observatory (acquired through the Gemini Science Archive), which is operated by the Association of Universities for Research in Astronomy, Inc., under a cooperative agreement 
with the NSF on behalf of the Gemini partnership: the National Science Foundation (United 
States), the Science and Technology Facilities Council (United Kingdom), the 
National Research Council (Canada), CONICYT (Chile), the Australian Research Council (Australia), 
Minist\'{e}rio da Ci\^{e}ncia, Tecnologia e Inova\c{c}$\tilde{a}$o (Brazil) 
and Ministerio de Ciencia, Tecnolog\'{i}a e Innovaci\'{o}n Productiva (Argentina).This research has also made use of the NASA/ IPAC Infrared Science Archive, which is operated by the Jet Propulsion Laboratory, California Institute of Technology, under contract with the National Aeronautics and Space Administration

\bibliography{Sabin_NGC5189_V3}

\label{lastpage}

\end{document}